\documentclass[conference]{IEEEtran}

\renewcommand\IEEEkeywordsname{Index Terms}
\usepackage{graphicx}
\usepackage{subcaption}
\usepackage{tikz,xparse}

\usetikzlibrary{patterns}

\usetikzlibrary{dsp,chains}
\usetikzlibrary{matrix}
\usepackage{mathptmx}
\usepackage{verbatim}
\usepackage{calc}
\usepackage{ifthen}
\usepackage{xifthen}
\usepackage{cancel}
\usepackage{bm}
\usepackage{verbatim}
\usepackage{multirow}
\usepackage{cite}
\usepackage[hyphens]{url}
\usepackage[nolist]{acronym} 
\usepackage{pgfplots}
\usetikzlibrary{arrows,shapes,graphs,graphs.standard,quotes,arrows.meta,decorations.markings,positioning}
\pgfplotsset{compat=newest}
\usepackage[bookmarks=false]{hyperref}
\usepackage{units}
\usepackage{amsmath, amsbsy, amssymb, latexsym }
\DeclareMathAlphabet{\mathcal}{OMS}{cmsy}{m}{n}
\hypersetup{bookmarksdepth=-2}
\usepackage{comment}
\usepackage[utf8]{inputenc}
\usepackage{xcolor}
\usepackage{enumitem}
\usepackage[linesnumbered,vlined, ruled]{algorithm2e}
\usepackage{algpseudocode}

\usepackage{marginnote}
\tikzset{>=latex}
\SetAlCapNameFnt{\footnotesize}
\SetAlCapFnt{\footnotesize}


\definecolor{mittelblau}{RGB}{0, 126, 198}
\definecolor{violettblau}{cmyk}{0.9, 0.6, 0, 0}
\definecolor{rot}{RGB}{238, 28 35}
\definecolor{apfelgruen}{RGB}{140, 198, 62}
\definecolor{gelb}{RGB}{255, 229, 0}
\definecolor{orange}{RGB}{244, 111, 33}
\definecolor{pink}{RGB}{237, 0, 140}
\definecolor{lila}{RGB}{128, 10, 145}
\definecolor{hellgrau}{RGB}{224, 224, 224}
\definecolor{mittelgrau}{RGB}{128, 128, 128}
\definecolor{dunkelgrau}{RGB}{80,80,80}
\definecolor{anthrazit}{RGB}{19, 31, 31}
\definecolor{darkgreen}{RGB}{34,139,34}
\definecolor{aqua}{RGB}{0, 255, 255}
\tikzset{
       vnd/.style={
        shape=circle,
        fill=black,
        draw,
        inner sep=0pt,
        minimum size=0.2cm},
        cnd/.style={
        shape=rectangle,
        fill=white,
        draw,
        minimum width=0.05mm,
        minimum height = 0.05mm}, 
         vndR/.style={
        shape=circle,
        fill=red,
        draw,
        inner sep=0pt,
        minimum size=0.2cm},
        cndR/.style={
        shape=rectangle,
        fill=white,
        draw=red,
        minimum width=0.05mm,
        minimum height = 0.05mm}
}

\IEEEoverridecommandlockouts

\begin{document}
	
\begin{NoHyper}
\title{Graph Search based Polar Code Design}

\author{\IEEEauthorblockN{Marvin Geiselhart, Andreas Zunker, Ahmed Elkelesh, Jannis Clausius and Stephan ten Brink}
	\IEEEauthorblockA{
		Institute of Telecommunications, Pfaffenwaldring 47, University of  Stuttgart, 70569 Stuttgart, Germany 
		\\\{geiselhart,elkelesh,clausius,tenbrink\}@inue.uni-stuttgart.de\\
	}
		\thanks{This work is supported by the German Federal Ministry of Education and Research (BMBF) within the project Open6GHub (grant no. 16KISK019).}
}

\maketitle

\begin{acronym}
\acro{ML}{maximum likelihood}
\acro{BP}{belief propagation}
\acro{BPL}{belief propagation list}
\acro{LDPC}{low-density parity-check}
\acro{BER}{bit error rate}
\acro{SNR}{signal-to-noise-ratio}
\acro{BPSK}{binary phase shift keying}
\acro{AWGN}{additive white Gaussian noise}
\acro{LLR}{Log-likelihood ratio}
\acro{MAP}{maximum a posteriori}
\acro{FER}{frame error rate}
\acro{BLER}{block error rate}
\acro{SCL}{successive cancellation list}
\acro{SC}{successive cancellation}
\acro{BI-DMC}{Binary Input Discrete Memoryless Channel}
\acro{CRC}{cyclic redundancy check}
\acro{CA-SCL}{CRC-aided successive cancellation list}
\acro{BEC}{Binary Erasure Channel}
\acro{BSC}{Binary Symmetric Channel}
\acro{BCH}{Bose-Chaudhuri-Hocquenghem}
\acro{RM}{Reed--Muller}
\acro{RS}{Reed-Solomon}
\acro{SISO}{soft-in/soft-out}
\acro{3GPP}{3rd Generation Partnership Project }
\acro{eMBB}{enhanced Mobile Broadband}
\acro{CN}{check node}
\acro{VN}{variable node}
\acro{GenAlg}{Genetic Algorithm}
\acro{CSI}{Channel State Information}
\acro{OSD}{ordered statistic decoding}
\acro{MWPC-BP}{minimum-weight parity-check BP}
\acro{FFG}{Forney-style factor graph}
\acro{MBBP}{multiple-bases belief propagation}
\acro{URLLC}{ultra-reliable low-latency communications}
\acro{DMC}{discrete memoryless channel}
\acro{SGD}{stochastic gradient descent}
\acro{QC}{quasi-cyclic}
\acro{NN}{neural network}
\acro{5G}{fifth generation mobile telecommunication}
\acro{SCAN}{soft cancellation}
\acro{AED}{automorphism ensemble decoding}
\acro{CCDF}{complementary cumulative distribution function}
\end{acronym}

\begin{abstract}
It is well known that to fulfill their full potential, the design of polar codes must be tailored to their intended decoding algorithm. While for \ac{SC} decoding, information theoretically optimal constructions are available, the code design for other decoding algorithms (such as \ac{BP} decoding) can only be optimized using extensive Monte Carlo simulations. 
We propose to view the design process of polar codes as a graph search problem and thereby approaching it more systematically. Based on this formalism, the  design-time complexity can be significantly reduced compared to state-of-the-art \ac{GenAlg} and deep learning-based design algorithms. Moreover, sequences of rate-compatible polar codes can be efficiently found. Finally, we analyze both the complexity of the proposed algorithm and the error-rate performance of the constructed codes.
\end{abstract}
\acresetall

\section{Introduction}

Polar codes, introduced by Arıkan, have attracted much interest due to their theoretical capability to achieve the capacity of the \ac{BI-DMC} under \ac{SC} decoding~\cite{ArikanMain} and their standardization in the \ac{5G}.
In the short blocklength regime, however, the performance of polar codes under \ac{SC} decoding is not satisfactory. Therefore, alternative decoding algorithms have been proposed to improve the error-rate performance (e.g., \ac{SCL} decoding \cite{talvardyList}, \ac{AED} \cite{polar_aed}), providing soft output (e.g., \ac{SCAN} decoding \cite{SCAN_Dec}) or reducing the latency (e.g., \ac{BP} decoding \cite{ArikanBP}). 
It has been shown that different channels and decoding algorithms require different code designs to achieve the best possible error-rate performance \cite{codeDesignGenAlg}.
While polar code design for \ac{BI-DMC} under \ac{SC} decoding is well studied, there exists no explicit construction optimized for other decoding algorithms. 
Consequently, finding suitable code designs for these decoders either is based on sub-optimal approximations, heuristics or requires extensive Monte Carlo simulations. 
In \cite{TUM_SCL_Construct}, codes for \ac{SCL} decoding are designed based on a heuristic, while in \cite{InfoTheorySCL}, designs are hand-crafted based on an information-theoretic analysis of the decoder.
Density evolution and its Gaussian approximation have been used in \cite{constructDE} and \cite{constructGaussian, GA, PiecewiseGA}, respectively, to design polar codes. For iterative \ac{BP} decoding, \ac{LLR} evolution has been proposed in~\cite{BP_LLR_Siegel}.
More generally applicable code design algorithms are based on Monte Carlo methods. In \cite{MC_BP_2}, the bitwise \ac{BER} is used to find reliable synthetic channels and successively generate the code design in a greedy fashion. 
A similar approach is used in \cite{Liu_unfreezing}, where the actual performance  of the codes is simulated instead of the \ac{BER}.

To allow for a broader search than greedy algorithms, the use of a \ac{GenAlg} has been proposed in \cite{codeDesignGenAlg}. Here, each code design is treated as an individual in a population that evolves over multiple generations using selection, crossover and mutation. Since then, the efficiency of \ac{GenAlg} has been improved by better crossover algorithms and caching \cite{EfficientGenAlg}.

Further, machine learning methods were applied to polar code design. In \cite{ebada2019deeplearningpolar}, the code design is learned via gradient descent through an unrolled \ac{BP} decoder. More recently, polar codes were learned via reinforcement learning \cite{PolarCodeRL}. In \cite{usingNeuralNetworks}, a \ac{NN} is trained to predict the \ac{FER} performance of polar code designs and then, a projected gradient algorithm is used to find the input to the \ac{NN} (i.e., a polar code design) that minimizes the \ac{FER}.

The main contributions of this paper can be summarized as follows:
\begin{itemize}
    \item We present a new perspective on polar code design as a problem on a graph
    \item First algorithms to optimize single code designs and rate-compatible reliability sequences are proposed
    \item We propose the use of confidence intervals as a general method to reduce the complexity of Monte Carlo simulation based code search. 
\end{itemize}

\section{Preliminaries}\label{sec:preliminaries}
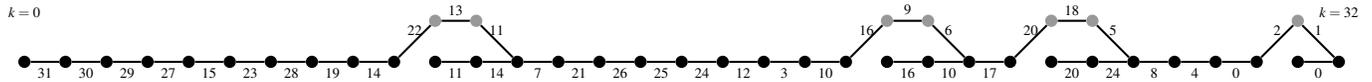
\begin{figure*}
    \centering
    \resizebox{\linewidth}{!}{\begin{tikzpicture}
\tikzstyle{code}=[circle,fill=black,inner sep=3pt,minimum size=1mm]
\tikzstyle{nocode}=[circle,inner sep=3pt,fill=black!40,minimum size=1mm]
\tikzstyle{basic} = [draw,minimum size=2mm,black]
\tikzstyle{edgenode} = [pos=.5,below]
\tikzstyle{edgenode_ab} = [pos=.5,above]
\tikzstyle{line} = [line width = 0.5mm]

\foreach \x in {0,1,...,32} {
    \node[code] (k\x) at (-\x,0) {};
}
\node[nocode] (k1u1) at (-1,1) {};
\node[nocode] (k6u5) at (-6,1) {};
\node[nocode] (k7u18) at (-7,1) {};
\node[nocode] (k10u6) at (-10,1) {};
\node[nocode] (k11u9) at (-11,1) {};
\node[nocode] (k21u11) at (-21,1) {};
\node[nocode] (k22u13) at (-22,1) {};

\node[above =of k32, yshift=-.2cm] {$k=0$};
\node[above =of k0, yshift=-.2cm] {$k=32$};

\draw[line] (k0) -- (k1) node[edgenode]  {0};
\draw[line] (k0) -- (k1u1) node[edgenode_ab]  {1};
\draw[line] (k1u1) -- (k2) node[edgenode_ab]  {2};
\draw[line] (k2) -- (k3) node[edgenode]  {0};
\draw[line] (k3) -- (k4) node[edgenode]  {4};
\draw[line] (k4) -- (k5) node[edgenode]  {8};
\draw[line] (k5) -- (k6) node[edgenode]  {24};
\draw[line] (k6) -- (k7) node[edgenode]  {20};
\draw[line] (k5) -- (k6u5) node[edgenode_ab]  {5};
\draw[line] (k6u5) -- (k7u18) node[edgenode_ab]  {18};
\draw[line] (k7u18) -- (k8) node[edgenode_ab]  {20};
\draw[line] (k8) -- (k9) node[edgenode]  {17};
\draw[line] (k9) -- (k10) node[edgenode]  {10};
\draw[line] (k10) -- (k11) node[edgenode]  {16};
\draw[line] (k9) -- (k10u6) node[edgenode_ab]  {6};
\draw[line] (k10u6) -- (k11u9) node[edgenode_ab]  {9};
\draw[line] (k11u9) -- (k12) node[edgenode_ab]  {16};

\draw[line] (k12) -- (k13) node[edgenode]  {10};
\draw[line] (k13) -- (k14) node[edgenode]  {3};
\draw[line] (k14) -- (k15) node[edgenode]  {12};
\draw[line] (k15) -- (k16) node[edgenode]  {24};
\draw[line] (k16) -- (k17) node[edgenode]  {25};
\draw[line] (k17) -- (k18) node[edgenode]  {26};
\draw[line] (k18) -- (k19) node[edgenode]  {21};
\draw[line] (k19) -- (k20) node[edgenode]  {7};
\draw[line] (k20) -- (k21) node[edgenode]  {14};
\draw[line] (k21) -- (k22) node[edgenode]  {11};

\draw[line] (k20) -- (k21u11) node[edgenode_ab]  {11};
\draw[line] (k21u11) -- (k22u13) node[edgenode_ab]  {13};
\draw[line] (k22u13) -- (k23) node[edgenode_ab]  {22};
\draw[line] (k23) -- (k24) node[edgenode]  {14};
\draw[line] (k24) -- (k25) node[edgenode]  {19};
\draw[line] (k25) -- (k26) node[edgenode]  {28};
\draw[line] (k26) -- (k27) node[edgenode]  {23};
\draw[line] (k27) -- (k28) node[edgenode]  {15};
\draw[line] (k28) -- (k29) node[edgenode]  {27};
\draw[line] (k29) -- (k30) node[edgenode]  {29};
\draw[line] (k30) -- (k31) node[edgenode]  {30};
\draw[line] (k31) -- (k32) node[edgenode]  {31};
\end{tikzpicture}}
    \caption{\footnotesize Sequence of $(N=32,k)$ polar codes for the \ac{AWGN} channel and \ac{BP} decoding. Black dots represent the best possible polar code for each code dimension~$k$, while gray dots are sub-optimal codes required to create a reliability sequence.}
    \label{fig:sequence32}
\end{figure*}

\subsection{Polar Codes}
Polar codes, as introduced in \cite{ArikanMain}, are based on the $n$-fold application of the basic \textit{channel transformation}, transforming $N=2^n$ identical channels into $N$ polarized synthetic channels. The subset $\mathcal{A}\subseteq \{0,...,N-1\}$ of synthetic channels with $|\mathcal{A}|=k$ is said to be reliable\footnote{The reliability refers to the information after decoding and, thus, is not a universal code property, but also dependent on the decoder.} and carries the information (i.e., information set), while the remaining $N-k$ synthetic channels $\mathcal{A}^c$ are said to be unreliable and thus transmit a frozen 0 (i.e., frozen set). 
The code $\mathcal C$ is defined by the encoding rule
\begin{equation*}
	\mathbf x = \mathbf u \cdot \mathbf G_N, \qquad \mathbf G_N = \begin{bmatrix} 1 & 0 \\ 1 & 1 \end{bmatrix}^{\otimes n},
\end{equation*}
with $\mathbf u_{\mathcal A} \in \{0,1\}^k \text{, } \mathbf u_{\mathcal A^c} = \mathbf 0$. Thus, the code rate is $R=\nicefrac{k}{N}$.
The choice of $\mathcal A$ is called \textit{polar code design} and optimal solutions are dependent on both the channel and the decoding algorithm~\cite{codeDesignGenAlg}.
An alternative notation for specifying the sets $\mathcal{A}$ and $\mathcal{A}^c$, respectively, is the binary vector $\mathbf{A}$ with
\begin{equation*}
A_i =     \begin{cases}
1 & \text{if } i \in \mathcal A\\
0 & \text{if } i \in \mathcal A^c
\end{cases}.
\end{equation*}
Throughout this paper, we will use $\mathcal{A}$-set, $\mathbf{A}$-vector and code~$\mathcal{C}$ notation interchangeably.

\subsection{Polar Code Reliability Sequences}
Practical applications require a simple change of the code rate whenever the channel conditions vary. For a fixed blocklength $N$, the code rate can be changed by moving some indices from $\mathcal{A}^c$ to $\mathcal{A}$ or vice-versa. A common way to specify the order of freezing/unfreezing is in form of a \textit{reliability sequence} $\mathbf{Q}$ that lists the indices of the synthetic channels in descending reliability order\footnote{In literature, ascending reliability is commonly used. However, descending order results in easier notation.}. To construct a polar code with a desired $k$, the $k$ most reliable (i.e., the first $k$) indices are chosen to be the information set, i.e.,
\begin{equation*}
\mathcal{A} = \{i \in Q_j | j < k\}.
\end{equation*}
Examples for reliability sequences are based on the Bhattacharyya parameter \cite{ArikanMain}, $\beta$-expansion \cite{BetaIngmard} and the 5G sequence \cite{polar5G2018}.

\textit{Remark:} Reliability sequences are in general sub-optimal. In other words, given a channel and decoding algorithm, the optimal code designs $\mathcal{A}_k$ for each $k$ do not necessarily fulfill $\mathcal{A}_{k-1} \subset \mathcal{A}_k$ and hence, do not necessarily form a sequence. Fig.~\ref{fig:sequence32} illustrates this property for $N=32$ and \ac{BP} decoding. Each black dot corresponds to the optimal code design for the respective code dimension $k$. There is no consecutive sequence of synthetic channels that contains all the best codes. Instead, for some code dimensions, sub-optimal codes (gray nodes) must be included to create a sequence.

\section{Polar Code Design on Graphs}

\subsection{Monte Carlo Simulation Based Code Search}
For most polar decoding algorithms besides \ac{SC} decoding, optimal explicit code constructions are unknown. Hence, one has to select good codes based on their measured performance. The performance is estimated at a pre-defined \ac{SNR} using Monte Carlo simulation.
With the number of simulated frame errors $N_\mathrm{FE}$ and trials $N_{\mathrm{T}}$, the accuracy of the simulation can be evaluated by a confidence interval $(P_\mathrm{FE,LB}, P_\mathrm{FE,UB})$ which contains the actual \ac{FER} $P_\mathrm{FE}$ of the code with a chosen probability $\gamma$, called the confidence level.
The frame errors are independent events, and hence, the number of observed frame errors $N_\mathrm{FE}$ is binomially distributed\footnote{In contrast, bit errors after decoding are not independent events, and hence, the outlined method only works for \ac{FER}.}. The confidence intervals can be thus computed using the relationship between binomial cumulative distribution and the incomplete beta function \cite{confidenceIntervals}.
However, according to the \emph{central limit theorem} for $N_\mathrm T \rightarrow \mathcal 1$, the distribution of the observed \ac{FER} $\hat P_\mathrm{FE} = N_\mathrm{FE}/N_\mathrm T $ approaches a normal distribution with mean $\mu = P_\mathrm{FE}$ and variance 
\begin{equation*}
	\sigma^2 = \frac{P_\mathrm{FE} \cdot \left(1 - P_\mathrm{FE} \right)}{N_\mathrm{T}}.
\end{equation*}
The confidence interval $(P_\mathrm{FE, LB}, P_\mathrm{FE,UB})$ of a Monte Carlo simulation can be approximated as ${(\hat P_\mathrm{FE,LB}, \hat P_\mathrm{FE,UB})} = {(\hat P_\mathrm{FE} - \delta, \hat P_\mathrm{FE} + \delta)}$ with
\begin{equation}
	\delta =\sqrt{\frac{\hat P_\mathrm{FE} \cdot \left(1 - \hat P_\mathrm{FE} \right)}{N_\mathrm{T}}} \cdot Q^{-1}(\alpha),\label{eq:confint}
\end{equation}
where $\alpha = \frac{1-\gamma}{2}$ and $Q^{-1}(\alpha)$ is the inverse of the \ac{CCDF} of the standard normal distribution \cite{confidenceIntervals}.
Note that the approximation becomes inaccurate if $N_\mathrm{FE}$ or $P_\mathrm{FE} \cdot N_\mathrm T$ are too small.
Confidence intervals can be used to compare two codes $\mathcal{C}_0$ and $\mathcal{C}_1$.
If ${P_{\mathrm{FE,UB}}(\mathcal{C}_0) < P_{\mathrm{FE,LB}}(\mathcal{C}_1)}$ holds, then the \ac{FER} of $\mathcal{C}_0$ is lower than that of $\mathcal{C}_1$ with probability
\begin{equation*}
    P\left[P_\mathrm{FE}(\mathcal{C}_0)<P_\mathrm{FE}(\mathcal{C}_1)\right] > 1 - \frac{(1-\gamma)^2}{4}. 
\end{equation*}
Furthermore, if an accurate estimation of the \acp{FER} is not required, the computational complexity can be reduced by terminating the Monte Carlo simulations as soon as it is determined which code is better.
Algorithm~\ref{alg:conf} generalizes this to finding the best $L$ codes of a set of codes ${\mathcal{L}=\{\mathcal{C}_0,\mathcal{C}_1,\dots\}}$.

\begin{algorithm}[tbh]
	\SetAlgoLined
	\LinesNumbered
	\SetKwInOut{Input}{Input}\SetKwInOut{Output}{Output}
	\Input{List $\mathcal{L}$ of codes $\mathcal{C}$, target number of codes $L$,\linebreak confidence level $\gamma$, $E_\mathrm{b}/N_0$}
	\Output{List $\mathcal{L}^*$ of $L$ best codes}
	
	$N_\mathrm{FE} \gets 0$\;
	$N_{\mathrm{T},\mathcal{C}} \gets 0 \quad \forall \mathcal{C} \in \mathcal{L} $ \;
	\While{$|\mathcal{L}| > L$}{
	    $N_\mathrm{FE} \gets N_\mathrm{FE} + 1$\;
	    \ForEach{$\mathcal{C} \in \mathcal{L}$}{
	        Simulate code $\mathcal{C}$ for 1 frame error, $N_\mathrm{T}$ trials at $E_\mathrm{b}/N_0$\;	        
	        $N_{\mathrm{T},\mathcal{C}} \gets N_{\mathrm{T},\mathcal{C}} + N_\mathrm{T}$\;
	        $\hat P_\mathrm{FE,\mathcal{C}} \gets N_\mathrm{FE}/N_{\mathrm{T},\mathcal{C}}$\;
	        compute $\hat P_\mathrm{FE,LB,\mathcal{C}}$ $\hat P_\mathrm{FE,UB,\mathcal{C}}$ from $ \gamma,\hat P_\mathrm{FE,\mathcal{C}},N_{\mathrm{T},\mathcal{C}}$ according to (\ref{eq:confint}) \;
	    }
	    $\hat P_\mathrm{FE,cutoff} \gets$ $L$-th smallest $\hat P_\mathrm{FE,UB,\mathcal{C}}$\;
	    $\mathcal{L} \gets \{\mathcal{C}\in \mathcal{L} \mid \hat P_\mathrm{FE,LB,\mathcal{C}} < \hat P_\mathrm{FE,cutoff}$\}\;
	}
    $\mathcal{L}^*\gets \mathcal{L}$\;
	
	\caption{\footnotesize Monte Carlo simulation based search of best $L$ code designs with early termination based on confidence intervals}\label{alg:conf}
\end{algorithm}

\subsection{The Graph of Polar Code Designs}
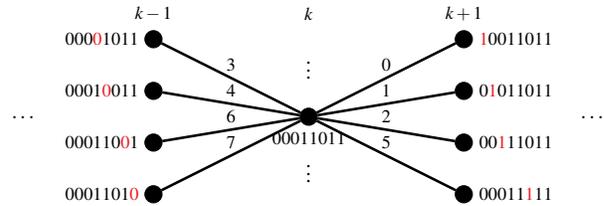
\begin{figure}
    \centering
\resizebox{.92\columnwidth}{!}{\begin{tikzpicture}
\tikzstyle{code}=[circle,draw,fill=black, font=\small]
\tikzstyle{basic} = [draw,minimum size=2mm,black]
\tikzstyle{edgenode} = [pos=.5,above]
\tikzstyle{line} = [line width = 0.5mm]

\node[] at (-3,2) {$k-1$};
\node[] at (0,2) {$k$};
\node[] at (3,2) {$k+1$};

\node[] at (-5.5,0) {\large$\dots$};
\node[] at (5.5,0) {\large$\dots$};

\node[] at (0,1) {\large$\vdots$};
\node[] at (0,-1) {\large$\vdots$};

\node[code, label=below:00011011] (mid) at (0,0) {};
\node[code, label= left:000{\color{rot}{0}}1011] (l1) at (-3,1.5) {};
\node[code, label= left:0001{\color{rot}{0}}011] (l2) at (-3,0.5) {};
\node[code, label= left:000110{\color{rot}{0}}1] (l3) at (-3,-0.5) {};
\node[code, label= left:0001101{\color{rot}{0}}] (l4) at (-3,-1.5) {};

\node[code, label= right:{\color{rot}{1}}0011011] (r1) at (3,1.5) {};
\node[code, label= right:0{\color{rot}{1}}011011] (r2) at (3,0.5) {};
\node[code, label= right:00{\color{rot}{1}}11011] (r3) at (3,-0.5) {};
\node[code, label= right:00011{\color{rot}{1}}11] (r4) at (3,-1.5) {};

\draw[line] (mid) -- (l1) node[edgenode]  {3};
\draw[line] (mid) -- (l2) node[edgenode]  {4};
\draw[line] (mid) -- (l3) node[edgenode]  {6};
\draw[line] (mid) -- (l4) node[edgenode]  {7};

\draw[line] (mid) -- (r1) node[edgenode]  {0};
\draw[line] (mid) -- (r2) node[edgenode]  {1};
\draw[line] (mid) -- (r3) node[edgenode]  {2};
\draw[line] (mid) -- (r4) node[edgenode]  {5};

\end{tikzpicture}}
    \caption{\footnotesize Excerpt of the code design graph for $N=8$.}
    \label{fig:graph_n8}
\end{figure}

\begin{figure}
    \centering
\resizebox{.85\columnwidth}{!}{\begin{tikzpicture}[yscale=0.5,xscale=1.1]
\tikzstyle{edgenode} = [pos=0.5,above=-0.6mm,font=\scriptsize]
\tikzstyle{edgenodex} = [pos=0.66,above=-0.6mm,font=\scriptsize]
\tikzstyle{edgenodey} = [pos=0.33,above=-0.6mm,font=\scriptsize]
\tikzstyle{code}=[circle,draw,fill=black, font=\small,inner sep=3pt]
\tikzstyle{line} = [line width = 0.33mm]

\node[code, label=above:0000]  (a0000) at (0.0,0.0) {};
\node[code, label=above:1000]  (a1000) at (2.0,3.0) {};
\node[code, label=above:0100]  (a0100) at (2.0,1.0) {};
\node[code, label=above:0010]  (a0010) at (2.0,-1.0) {};
\node[code, label=above:0001]  (a0001) at (2.0,-3.0) {};
\node[code, label=above:1100]  (a1100) at (4.0,5.0) {};
\node[code, label=above:1010]  (a1010) at (4.0,3.0) {};
\node[code, label=above:1001]  (a1001) at (4.0,1.0) {};
\node[code, label=above:0110]  (a0110) at (4.0,-1.0) {};
\node[code, label=above:0101]  (a0101) at (4.0,-3.0) {};
\node[code, label=above:0011]  (a0011) at (4.0,-5.0) {};
\node[code, label=above:1110]  (a1110) at (6.0,3.0) {};
\node[code, label=above:1101]  (a1101) at (6.0,1.0) {};
\node[code, label=above:1011]  (a1011) at (6.0,-1.0) {};
\node[code, label=above:0111]  (a0111) at (6.0,-3.0) {};
\node[code, label=above:1111]  (a1111) at (8.0,0.0) {};

\draw[line] (a0000) -- (a1000) node[edgenode]  {0};
\draw[line] (a0000) -- (a0100) node[edgenode]  {1};
\draw[line] (a0000) -- (a0010) node[edgenode]  {2};
\draw[line] (a0000) -- (a0001) node[edgenode]  {3};
\draw[line] (a1000) -- (a1100) node[edgenode]  {1};
\draw[line] (a1000) -- (a1010) node[edgenode]  {2};
\draw[line] (a1000) -- (a1001) node[edgenode]  {3};
\draw[line] (a0100) -- (a1100) node[edgenodex]  {0};
\draw[line] (a0100) -- (a0110) node[edgenode]  {2};
\draw[line] (a0100) -- (a0101) node[edgenodey]  {3};
\draw[line] (a0010) -- (a1010) node[edgenode]  {0};
\draw[line] (a0010) -- (a0110) node[edgenodex]  {1};
\draw[line] (a0010) -- (a0011) node[edgenodey]  {3};
\draw[line] (a0001) -- (a1001) node[edgenodey]  {0};
\draw[line] (a0001) -- (a0101) node[edgenodex]  {1};
\draw[line] (a0001) -- (a0011) node[edgenode]  {2};
\draw[line] (a1100) -- (a1110) node[edgenode]  {2};
\draw[line] (a1100) -- (a1101) node[edgenodey]  {3};
\draw[line] (a1010) -- (a1110) node[edgenodex]  {1};
\draw[line] (a1010) -- (a1011) node[edgenodey]  {3};
\draw[line] (a1001) -- (a1101) node[edgenodey]  {1};
\draw[line] (a1001) -- (a1011) node[edgenode]  {2};
\draw[line] (a0110) -- (a1110) node[edgenodex]  {0};
\draw[line] (a0110) -- (a0111) node[edgenode]  {3};
\draw[line] (a0101) -- (a1101) node[edgenode]  {0};
\draw[line] (a0101) -- (a0111) node[edgenodey]  {2};
\draw[line] (a0011) -- (a1011) node[edgenodey]  {0};
\draw[line] (a0011) -- (a0111) node[edgenode]  {1};
\draw[line] (a1110) -- (a1111) node[edgenode]  {3};
\draw[line] (a1101) -- (a1111) node[edgenode]  {2};
\draw[line] (a1011) -- (a1111) node[edgenode]  {1};
\draw[line] (a0111) -- (a1111) node[edgenode]  {0};
  
\end{tikzpicture}}
    \caption{\footnotesize Complete graph with all polar code designs of length $N=4$.}
    \label{fig:graph_n4}
\end{figure}
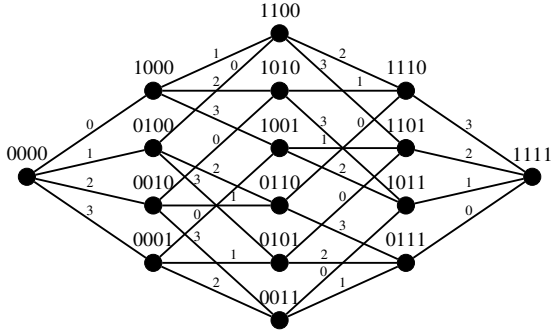

To relate different polar code designs to each other, we propose to use a (directed) graph. Each polar code design (i.e., $\mathbf{A}$-vector) corresponds to a vertex. Two codes $\mathcal{A}$ and $\mathcal{A}'$ differing exactly by one frozen/unfrozen bit are connected by an edge, and the edge label indicates the bit position in which they differ, i.e.,
\begin{equation*}
    \mathcal{A} \frac{j}{\qquad} \mathcal{A}' \quad \Leftrightarrow \quad \mathcal{A}' = \mathcal{A} \cup \{j\}.
\end{equation*}
Note that this is identical to the Hasse diagram of all information sets ordered by inclusion. We define the partial order
\begin{equation*}
    \mathcal{A} \prec \mathcal{A}' \quad \Leftrightarrow \quad \mathcal{A} \subset \mathcal{A}'
\end{equation*}
that can also compare codes not directly neighboring, but connected via a chain of edges. 
This notion of order is motivated by the fact that the \acp{FER} of two codes $\mathcal{A}$ and $\mathcal{A}'$ with $\mathcal{A} \prec \mathcal{A}'$ at identical $E_\mathrm{s}/N_0$ fulfill $P_\mathrm{FE}(\mathcal{A}) \le P_\mathrm{FE}(\mathcal{A}')$, as the decoder of $\mathcal{A}$ has access to more a priori information (additional frozen bits) than the decoder of $\mathcal{A}'$. Therefore, the graph implies some local ``smoothness'' of the \ac{FER} in the neighborhood around each code.

In Fig. \ref{fig:graph_n8} an excerpt of the graph for $N=8$ is shown. Note that we implicitly assume increasing code dimensions from left to right and, thus, the direction of the edges is omitted for readability. 
Fig. \ref{fig:graph_n4} shows the complete graph for all polar codes with blocklength $N=4$. 

\subsection{Optimization of a Single Polar Code Design}
\begin{figure}
    \centering
\resizebox{.85\columnwidth}{!}{\begin{tikzpicture} 
\tikzstyle{code}=[circle,draw,fill=black, inner sep=3pt]
\tikzstyle{basic} = [draw,minimum size=2mm,black]
\tikzstyle{edgenode} = [pos=.425, above, inner sep=2pt]
\tikzstyle{line} = [line width = 0.5mm]

\node[] at (0,0.5) {$k$};

\node[code, label=right:\colorbox{apfelgruen!50}{00011011}] (mid) at (0,0) {};
\node[] at (2.5,0) {1.};
\node[] at (-7.5,-1) {2.};
\node[] at (2.5,-4.5) {3.};

\node[] at (-5,0.5) {$k-1$};

\node[code, label= left:000{\color{rot}{0}}1011~~] (l1) at (-5,0) {};
\node[code, label= left:\colorbox{apfelgruen!50}{0001{\color{rot}{0}}011}] (l2) at (-5,-1) {};
\node[code, label= left:000110{\color{rot}{0}}1~~] (l3) at (-5,-2) {};
\node[code, label= left:0001101{\color{rot}{0}}~~] (l4) at (-5,-3) {};

\draw[line] (mid) -- (l1) node[edgenode]  {3};
\draw[line] (mid) -- (l2) node[edgenode]  {4};
\draw[line] (mid) -- (l3) node[edgenode]  {6};
\draw[line] (mid) -- (l4) node[edgenode]  {7};

\node[code, label= right:~{\color{rot}{1}}001{\color{rot}{0}}011] (r1) at (0,-1.5) {};
\node[code, label= right:~0{\color{rot}{1}}01{\color{rot}{0}}011] (r2) at (0,-2.5) {};
\node[code, label= right:~00{\color{rot}{1}}1{\color{rot}{0}}011] (r3) at (0,-3.5) {};
\node[code, label= right:\colorbox{apfelgruen!50}{0001{\color{rot}{01}}11}] (r4) at (0,-4.5) {};

\draw[line] (l2) -- (r1) node[edgenode]  {0};
\draw[line] (l2) -- (r2) node[edgenode]  {1};
\draw[line] (l2) -- (r3) node[edgenode]  {2};
\draw[line] (l2) -- (r4) node[edgenode]  {5};

\end{tikzpicture}}
    \caption{\footnotesize Example of graph search for a single polar code design, $N=8$.}
    \label{fig:single_code_design}
\end{figure}
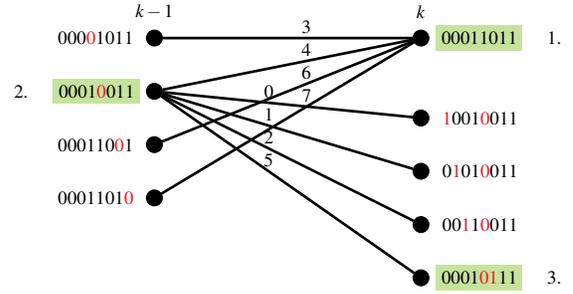
A first algorithm to traverse the graph in order to find an optimized, single code design is the \textit{bit swapping} algorithm shown in Algorithm~\ref{alg:single}. Starting from any code design $\mathcal{C}_0$ with the desired code dimension $k$ (e.g., using $\beta$-expansion), information and frozen bits are alternately exchanged. The algorithm keeps a list of the $L$ best candidates and estimates the performance of its left neighbors using Algorithm~\ref{alg:conf}. Then, the right neighbors of these codes are simulated. This way, the algorithm ``zig-zags'' through the graph between $k$ and $k-1$, until no more progress is made. An example for a single iteration of the algorithm is illustrated in Fig. \ref{fig:single_code_design} for $N=8$.

\begin{algorithm}[tbh]
    \caption{\footnotesize Optimization of a single code design}\label{alg:single}
	\SetAlgoLined
	\LinesNumbered
	\SetKwInOut{Input}{Input}\SetKwInOut{Output}{Output}
	\Input{Start code $\mathcal{C}_0$, list size $L$, confidence level $\gamma$, $E_\mathrm{b}/N_0$}
	\Output{List $\mathcal{L}^*$ of $L$ best codes}
	$\mathcal{L} \gets \{ \mathcal{C}_0\}$\;
	\While{no further improvement}{
	    $\mathcal{L} \gets \bigcup_{\mathcal{C}\in \mathcal{L}} (\text{left neighbors of } \mathcal{C}$)\;
	    $\mathcal{L} \gets \operatorname{Algorithm~1}(\mathcal{L}, L, \gamma, E_\mathrm{b}/N_0)$\;
	    $\mathcal{L} \gets \bigcup_{\mathcal{C}\in \mathcal{L}} (\text{right neighbors of } \mathcal{C})$\;
	    $\mathcal{L} \gets \operatorname{Algorithm~1}(\mathcal{L}, L, \gamma, E_\mathrm{b}/N_0)$\;
	}
	$\mathcal{L}^*\gets \mathcal{L}$\;
	
\end{algorithm}

\subsection{Optimizing a Bit Reliability Sequence}
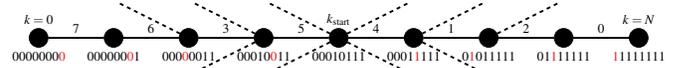
\begin{figure}
    \centering
\resizebox{\columnwidth}{!}{\begin{tikzpicture}
\tikzstyle{code}=[circle,draw,fill=black, inner sep=5pt]
\tikzstyle{basic} = [draw,minimum size=2mm,black]
\tikzstyle{edgenode} = [pos=.5,above, inner sep= 3pt]
\tikzstyle{line} = [line width = 0.5mm]

\node[] at (0,0.5) {$k_\mathrm{start}$};

\node[code, label=below:00010111] (k4) at (0,0) {};

\node[code, label=below:0001{\color{rot}{1}}111] (k5) at (2,0) {};
\node[code, label=below:00010{\color{rot}{0}}11] (k3) at (-2,0) {};
\node (d51) at (2,-1) {};
\node (d31) at (-2,-1) {};
\node (d52) at (2,1) {};
\node (d32) at (-2,1) {};

\draw[line] (k4) -- (k3) node[edgenode]  {5};
\draw[line] (k4) -- (k5) node[edgenode]  {4};
\draw[line,dashed] (k4) -- (d31) {};
\draw[line,dashed] (k4) -- (d51) {};
\draw[line,dashed] (k4) -- (d32) {};
\draw[line,dashed] (k4) -- (d52) {};

\node[code, label=below:0{\color{rot}{1}}011111] (k6) at (4,0) {};
\node[code, label=below:000{\color{rot}{0}}0011] (k2) at (-4,0) {};
\draw[line] (k3) -- (k2) node[edgenode]  {3};
\draw[line] (k5) -- (k6) node[edgenode]  {1};

\node (d61) at (4,-1) {};
\node (d21) at (-4,-1) {};
\node (d62) at (4,1) {};
\node (d22) at (-4,1) {};
\draw[line,dashed] (k3) -- (d21) {};
\draw[line,dashed] (k5) -- (d61) {};
\draw[line,dashed] (k3) -- (d22) {};
\draw[line,dashed] (k5) -- (d62) {};

\node[code, label=below:01{\color{rot}{1}}11111] (k7) at (6,0) {};
\node[code, label=below:000000{\color{rot}{0}}1] (k1) at (-6,0) {};
\draw[line] (k2) -- (k1) node[edgenode]  {6};
\draw[line] (k6) -- (k7) node[edgenode]  {2};

\node (d71) at (6,1) {};
\node (d11) at (-6,1) {};
\draw[line,dashed] (k2) -- (d11) {};
\draw[line,dashed] (k6) -- (d71) {};

\node[] at (-8,0.5) {$k=0$};
\node[] at (8,0.5) {$k=N$};
\node[code, label=below:{\color{rot}{1}}1111111] (k8) at (8,0) {};
\node[code, label=below:0000000{\color{rot}{0}}] (k0) at (-8,0) {};
\draw[line] (k1) -- (k0) node[edgenode]  {7};
\draw[line] (k7) -- (k8) node[edgenode]  {0};

\end{tikzpicture}}
    \caption{\footnotesize Example of graph search for a rate-compatible sequence, $N=8$.}
    \label{fig:sequence}
\end{figure}

A similar approach to Algorithm~\ref{alg:single} can be used to optimize a rate-compatible sequence of codes. This procedure is listed in Algorithm~\ref{alg:sequence}. Starting from a list of good codes that was found using Algorithm~\ref{alg:single} for some starting code dimension $k_\mathrm{start}$, the algorithm develops sequences of neighboring codes outwards to $k=0$ and $k=N$. In each step, the best $L$ sequences $S$ are kept based on a path metric
\begin{equation}
    \tau(S) = \sum_{\mathcal{C} \in S} \log\frac{P_\mathrm{FE}(\mathcal{C})}{P_{\mathrm{FE,best},k(\mathcal{C})}} = \sum_{\mathcal{C} \in S} \log P_\mathrm{FE}(\mathcal{C}) + c, \label{eq:pathmetric}
\end{equation}
where $P_{\mathrm{FE,best},k(\mathcal{C})}$ is the \ac{FER} of the best found code for the same code dimension as $\mathcal{C}$. This path metric can be interpreted as the error-rate loss of the codes in the sequence versus the best codes that are possible for each $k$. 
This way, the algorithm aims at finding a good compromise of decently performing codes under the constraint that they form a sequence.
This constraint is enforced by lines \ref{algline:augment1} and \ref{algline:augment2}, where the currently found paths are augmented by appending (or pre-pending, respectively) only neighboring codes in the currently simulated batch $\mathcal{L}_k$. If multiple codes neighbor the last code $S_\mathrm{last}$ (or first code $S_\mathrm{first}$) in the sequence $S$, the sequence is duplicated for each option. Likewise, the work-list of codes to simulate in the next step includes all codes neighboring $S_\mathrm{last}$ (line \ref{algline:moveright}) and $S_\mathrm{first}$ (line \ref{algline:moveleft}), respectively.
For a list size of $L=1$ and starting code dimension $k_\mathrm{start}=0$, the algorithm degenerates to the greedy procedure presented in \cite{Liu_unfreezing}. 
Fig. \ref{fig:sequence} illustrates Algorithm~\ref{alg:sequence} for $N=8$ and $k_\mathrm{start}=4$. The bit reliability sequence $\mathbf Q$ can be extracted as the sequence of edge labels on the path from $k=0$ to $k=N$; in this example $\mathbf Q = [7,6,3,5,4,1,2,0]$.

\begin{algorithm}[tbh]
	\SetAlgoLined
	\LinesNumbered
	\SetKwInOut{Input}{Input}\SetKwInOut{Output}{Output}
	\Input{List of start codes $\mathcal{L}_{k_\mathrm{start}}$, $k_\mathrm{start}$, list size $L$, confidence level $\gamma$, $E_\mathrm{b}/N_0$}
	\Output{Best sequence $S^*$}
	$k_\mathrm{min}\gets k_\mathrm{start}, \,k_\mathrm{max}\gets k_\mathrm{start}$\;
	$\mathcal{L}_{k_\mathrm{start}} \gets \operatorname{Algorithm~1}(\mathcal{L}_{k_\mathrm{start}}, L, \gamma, E_\mathrm{b}/N_0)$\;
    $\mathcal{S}_\mathrm{paths} \gets \{ [\mathcal{C}] \mid \mathcal{C} \in \mathcal{L}_k \}$\;
	
	\While{$k_\mathrm{min}>0$ or $k_\mathrm{max}<N$}{
	    \If {$k_\mathrm{max}<N$}{
	        $k_\mathrm{max} \gets k_\mathrm{max}+1$\;
	        $\mathcal{L}_{k_\mathrm{max}} \gets \bigcup_{S\in \mathcal{S}_\mathrm{paths}} (\text{right neighbors of } S_\mathrm{last}$)\;  \label{algline:moveright}
	        $\mathcal{L}_{k_\mathrm{max}} \gets \operatorname{Algorithm~1}(\mathcal{L}_{k_\mathrm{max}}, L, \gamma, E_\mathrm{b}/N_0)$\;
	        Augment $S \in \mathcal{S}_\mathrm{paths}$ using codes $\mathcal{C} \in \mathcal{L}_{k_\mathrm{max}}$\; \label{algline:augment1}
	    }
	    \If {$k_\mathrm{min}>0$}{
	        $k_\mathrm{min}\gets k_\mathrm{min}-1$\;
	        $\mathcal{L}_{k_\mathrm{min}} \gets \bigcup_{S\in \mathcal{S}_\mathrm{paths}} (\text{left neighbors of } S_\mathrm{first}$)\;
	        \label{algline:moveleft}
	        $\mathcal{L}_{k_\mathrm{min}} \gets \operatorname{Algorithm~1}(\mathcal{L}_{k_\mathrm{min}}, L, \gamma, E_\mathrm{b}/N_0)$\;
	        Augment $S \in \mathcal{S}_\mathrm{paths}$ using codes $\mathcal{C} \in \mathcal{L}_{k_\mathrm{min}}$\; \label{algline:augment2}
	    }
	    Prune $\mathcal{S}_\mathrm{paths}$ to best $L$ paths w.r.t. to $\tau(S)$ from  (\ref{eq:pathmetric})\;
	}
	$S^* \gets \arg\min_{S\in \mathcal{S}_\mathrm{paths}} \tau(S)$\;
	\caption{\footnotesize Rate-compatible polar code sequence optimization}\label{alg:sequence}
\end{algorithm}

\section{Results}

\subsection{Single Code Design Optimization}
\begin{figure}
\centering
\resizebox{\columnwidth}{!}{\input{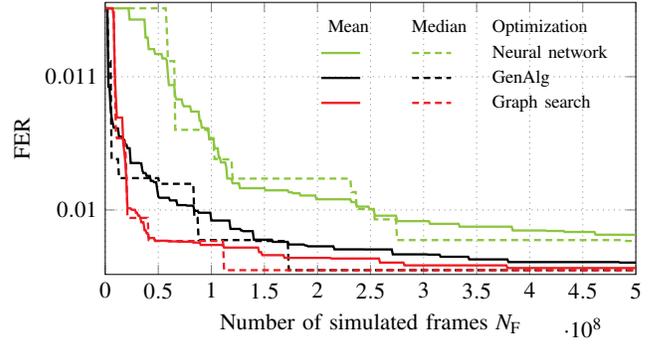}}
\caption{\footnotesize Design complexity in terms of simulation effort vs. the achievable \ac{FER}. The lines record the mean and median of 11 independent optimization runs for each optimizer.}
\label{fig:complexity}
\end{figure}	

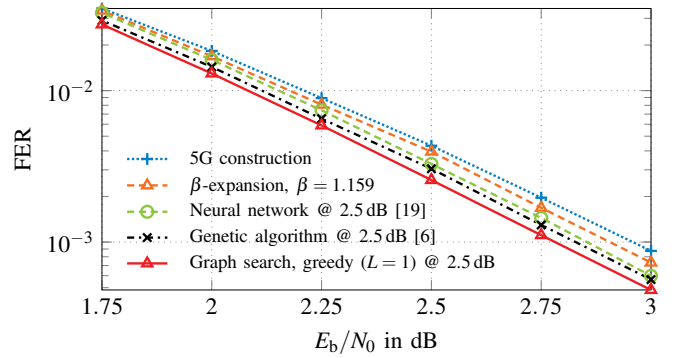
\begin{figure}
	\centering
	\resizebox{\columnwidth}{!}{\begin{tikzpicture}
\begin{axis}[
	width=\linewidth,
	height=.6\linewidth,
	scale=1.125,
	grid style={dotted,gray},
	xmajorgrids,
	yminorticks=true,
	ymajorgrids,
	legend columns=1,
	legend pos=south west,   
	legend cell align={left},
	legend style={fill=none,text opacity=1, draw=none},
	xtick={0,0.25,...,3.5},
	xlabel={$E_\mathrm b/N_0$ in dB},
	ylabel={FER},
	legend image post style={mark indices={}},
	ymode=log,
	mark size=1.5pt,
	xmin=1.75,
	xmax=3,
	ymin=0.0004830033534922833,
	ymax=0.03485261699587868 
]

\addplot [color=mittelblau,line width = 1pt,mark=+,mark size=2.5pt,mark options={solid},densely dotted]
table[col sep=comma]{
0.0,0.6732873253661
0.25,0.5444399074452158
0.5,0.4086636697997548
0.75,0.28803917332757256
1.0,0.1896993265673907
1.25,0.11397634990739422
1.5,0.06657567990413102
1.75,0.03485261699587868
2.0,0.01828462765641356
2.25,0.008921342751297497
2.5,0.004325137917835355
2.75,0.001963271123815657
3.0,0.0008742498662944111
3.25,0.00036095017245296864
3.5,0.00015054461207043366
};
\label{plot:RCA_optimized}
\addlegendentry{\footnotesize 5G construction};

\addplot [color=orange,line width=1pt,mark=triangle,mark size=2.5pt,mark options={solid},densely dashed]
table[col sep=comma]{
0.0,0.6825938566552902
0.25,0.5542469170015242
0.5,0.4142073107590349
0.75,0.2932121389825539
1.0,0.19015021867275148
1.25,0.11109259567849802
1.5,0.06428284451586982
1.75,0.03377579626439693
2.0,0.016869593822354742
2.25,0.008079175924055746
2.5,0.0039543978836062525
2.75,0.0016833918662713502
3.0,0.000732435465111169
3.25,0.00030913647440207015
3.5,0.00012329251806443445
};
\label{plot:beta}
\addlegendentry{\footnotesize $\beta$-expansion, $\beta = 1.159$};

\addplot [color=apfelgruen,line width = 1pt,mark=o,mark size=2.5pt,mark options={solid},densely dashed]
table[col sep=comma, row sep=crcr]{
0.0,0.6926406926406926\\
0.25,0.5621135469364812\\
0.5,0.42025635637739023\\
0.75,0.29357798165137616\\
1.0,0.19071231047964146\\
1.25,0.11041488392635328\\
1.5,0.061723632435768845\\
1.75,0.032621105855488504\\
2.0,0.015961819328167023\\
2.25,0.007412390180806728\\
2.5,0.003281109014847018\\
2.75,0.0014444052047697148\\
3.0,0.0005984603410805022\\
3.25,0.0002455236737609095\\
};
\label{plot:beta1.189}
\addlegendentry{\footnotesize Neural network @ \unit[2.5]{dB} \cite{usingNeuralNetworks} };

\addplot [color=black,line width = 1pt,mark=x,mark size=2.5pt,mark options={solid},dashdotted]
table[col sep=comma]{
0.0,0.6759040216289287
0.25,0.5360493165371214
0.5,0.402576489533011
0.75,0.2819283901888920
1.0,0.1786192730195588
1.25,0.1047559187094070
1.5,0.05617898625019311
1.75,0.029042329194801424
2.0,0.014335838521114898
2.25,0.006521391795436982
2.5,0.003048940836065311
2.75,0.001296854964603965
3.0,0.000565667250482408
3.25,0.0002556131856783766
3.5,0.00010835772036702819
};
\label{plot:beta1.133}
\addlegendentry{\footnotesize Genetic algorithm @ \unit[2.5]{dB} \cite{codeDesignGenAlg}};

\addplot [color=rot,line width = 1pt,mark=triangle,mark size=2.5pt,mark options={solid},solid]
table[col sep=comma, row sep=crcr]{
0.0,0.6685609226140732\\
0.25,0.5401026194977046\\
0.5,0.39844606036457814\\
0.75,0.27799013135033707\\
1.0,0.1763279700242451\\
1.25,0.1013993104846887\\
1.5,0.054381075385765754\\
1.75,0.027349492325048717\\
2.0,0.012970210668646785\\
2.25,0.005887045262547501\\
2.5,0.002568477207654319\\
2.75,0.0011105450416246164\\
3.0,0.0004830033534922833\\
};
\label{plot:beta1.133}
\addlegendentry{\footnotesize Graph search, greedy ($L=1$) @ \unit[2.5]{dB}};

\end{axis}
\end{tikzpicture}}
	\caption{\footnotesize Performance of (512,128) polar codes under BP decoding with $N_\mathrm{it,max}~=~20$ iterations.}
	\label{fig:fer512}
\end{figure}

We evaluate Algorithm~\ref{alg:single} for designing polar codes for the \ac{AWGN} channel and \ac{BP} decoding. For more information on \ac{BP}, we refer the interested reader to \cite{ISWCS_Error_Floor}. We compare the proposed method to optimizations using the deep learning approach from \cite{usingNeuralNetworks} and the \ac{GenAlg} proposed in \cite{codeDesignGenAlg} with the complexity reduction improvements from \cite{EfficientGenAlg}.
For the deep learning based method, the \ac{NN} consists of three dense layers with 128 neurons each and it is trained for 100 epochs per design algorithm iteration.
The \ac{GenAlg} uses a population size of 50. 

First, we design $(128,64)$ polar codes for $N_\mathrm{it,max}=100$ \ac{BP} decoding iterations at an \ac{SNR} $E_\mathrm{b}/N_0~=~3~\text{dB}$. The graph search algorithm uses a list size $L=4$ and $\gamma=0.8$. We notice that all algorithms converge to the identical, presumably globally optimal code design with the same \ac{FER} performance.

Therefore, to compare the algorithms quantitatively, we record the total number of frames transmitted in the Monte Carlo simulation. As all algorithms are incremental and intermediate solutions can be taken at any step in the optimization progress, we plot the mean and median \ac{FER} performance of the best codes from 11 independent runs of each optimizer in Fig. \ref{fig:complexity}. We can see that the \ac{NN}-based method has the largest design complexity as it requires a large data-set until the projected gradient method can start to produce gains. The \ac{GenAlg} starts off the fastest, however, then converges more slowly than the proposed graph search, which needs the least complexity to reliably converge to the optimal code design.

Next, we design longer (512,128) codes for $N_\mathrm{it,max}=20$ \ac{BP} iterations at $E_\mathrm{b}/N_0~=~2.5~\text{dB}$. We compare the three Monte Carlo based designs and also the 5G design as well as the $\beta$-expansion based design with an optimized value for $\beta=1.159$ in Fig. \ref{fig:fer512}. Here, the Monte Carlo optimized code designs perform better than the standardized codes and $\beta$-expansion. Moreover, the graph search designed a code outperforming also the \ac{GenAlg}, even without a list (i.e., $L=1$).

\subsection{Bit Reliability Sequence}
\begin{figure}
    \centering
\resizebox{\columnwidth}{!}{\begin{tikzpicture}
\begin{axis}[
	width=\linewidth,
	height=.6\linewidth,
	scale=1.125,
	grid style={dotted,gray},
	xmajorgrids,
	yminorticks=true,
	ymajorgrids,
	legend columns=1,
	legend style={at={(0.4,0.97)},anchor=north,draw=none,fill=none},
	legend cell align={left},
	xtick={0,16,...,128},
	xlabel={Code dimension $k$},
	ylabel={$E_\mathrm{b}/N_0$ required for $P_\mathrm{FE}\le 10^{-3}$},
	legend image post style={mark indices={}},
	mark size=1.5pt,
	xmin=0,
	xmax=128,
	ymin=3.5,
	ymax=6,
]

\addplot [color=orange,line width = 1pt,densely dashed]
table[col sep=comma]{
1,6.764
2,7.330
3,6.280
4,5.821
5,5.414
6,5.029
7,5.472
8,5.099
9,5.053
10,5.050
11,4.979
12,4.732
13,4.631
14,4.572
15,4.489
16,4.327
17,4.330
18,4.120
19,4.062
20,4.544
21,4.295
22,4.145
23,3.984
24,4.216
25,4.090
26,3.954
27,4.127
28,4.279
29,4.144
30,4.148
31,4.016
32,4.117
33,3.962
34,4.027
35,4.051
36,4.069
37,4.056
38,3.974
39,3.961
40,3.937
41,3.911
42,3.867
43,3.879
44,3.779
45,3.762
46,3.710
47,4.159
48,4.097
49,4.044
50,3.970
51,3.913
52,3.862
53,4.106
54,4.058
55,4.000
56,3.948
57,4.211
58,4.137
59,4.367
60,4.303
61,4.237
62,4.334
63,4.261
64,4.186
65,4.321
66,4.376
67,4.314
68,4.377
69,4.469
70,4.369
71,4.457
72,4.389
73,4.431
74,4.445
75,4.485
76,4.420
77,4.450
78,4.469
79,4.466
80,4.518
81,4.513
82,4.451
83,4.451
84,4.474
85,4.936
86,4.901
87,4.864
88,4.853
89,4.823
90,4.794
91,5.103
92,5.072
93,5.017
94,4.986
95,4.967
96,5.248
97,5.221
98,5.502
99,5.429
100,5.558
101,5.503
102,5.460
103,5.614
104,5.685
105,5.648
106,5.740
107,5.814
108,5.918
109,5.863
110,5.929
111,5.981
112,6.056
113,6.127
114,6.171
115,6.225
116,6.279
117,6.824
118,6.842
119,6.829
120,6.840
121,7.148
122,7.168
123,7.398
124,7.642
125,7.849
126,8.050
127,8.212
};
\label{plot:beta1.18}
\addlegendentry{\footnotesize $\beta$-expansion, $\beta=2^{\nicefrac{1}{4}}$};

\addplot [color=mittelblau,line width = 1pt,dashdotted]
table[col sep=comma]{
1,6.750
2,7.305
3,6.315
4,5.834
5,5.406
6,5.029
7,4.940
8,5.099
9,5.018
10,4.886
11,4.819
12,4.690
13,4.652
14,4.547
15,4.485
16,4.341
17,4.329
18,4.130
19,4.087
20,3.929
21,3.854
22,4.185
23,4.012
24,3.921
25,4.083
26,3.976
27,4.145
28,4.014
29,4.148
30,4.006
31,4.019
32,3.945
33,3.976
34,3.888
35,3.907
36,3.961
37,3.971
38,3.952
39,3.849
40,3.856
41,3.870
42,3.820
43,3.780
44,3.771
45,3.777
46,3.727
47,3.694
48,4.075
49,4.060
50,3.975
51,3.938
52,3.864
53,3.821
54,4.072
55,4.004
56,3.925
57,3.915
58,4.146
59,4.099
60,4.033
61,3.963
62,4.178
63,4.235
64,4.195
65,4.136
66,4.268
67,4.207
68,4.276
69,4.208
70,4.296
71,4.332
72,4.408
73,4.343
74,4.392
75,4.346
76,4.327
77,4.370
78,4.389
79,4.386
80,4.384
81,4.483
82,4.458
83,4.453
84,4.459
85,4.452
86,4.443
87,4.515
88,4.844
89,4.825
90,4.795
91,4.763
92,4.733
93,5.023
94,4.982
95,4.965
96,4.950
97,5.198
98,5.162
99,5.435
100,5.386
101,5.354
102,5.446
103,5.616
104,5.571
105,5.659
106,5.618
107,5.757
108,5.858
109,5.936
110,5.980
111,5.982
112,6.059
113,6.158
114,6.188
115,6.268
116,6.282
117,6.430
118,6.469
119,6.840
120,6.859
121,6.927
122,7.164
123,7.407
124,7.641
125,7.863
126,8.045
127,8.217
};
\label{plot:5gseq}
\addlegendentry{\footnotesize 5G construction};

\addplot [color=black,line width = 1pt,densely dotted]
table[col sep=comma]{
1,6.764
2,7.346
3,6.280
4,5.834
5,5.669
6,5.636
7,5.571
8,5.847
9,5.945
10,5.471
11,5.320
12,5.085
13,5.070
14,4.885
15,4.871
16,4.832
17,4.784
18,4.634
19,4.575
20,4.519
21,4.414
22,4.268
23,4.228
24,4.150
25,4.209
26,4.153
27,4.081
28,3.979
29,3.925
30,3.920
31,3.891
32,3.869
33,3.825
34,3.826
35,3.798
36,3.750
37,3.749
38,3.726
39,3.708
40,3.724
41,3.666
42,3.666
43,3.628
44,3.604
45,3.600
46,3.621
47,3.613
48,3.633
49,3.657
50,3.630
51,3.632
52,3.642
53,3.666
54,3.674
55,3.680
56,3.677
57,3.659
58,3.664
59,3.681
60,3.677
61,3.681
62,3.742
63,3.800
64,3.837
65,3.887
66,3.899
67,3.934
68,3.980
69,4.066
70,4.102
71,4.094
72,4.100
73,4.201
74,4.245
75,4.285
76,4.257
77,4.304
78,4.325
79,4.336
80,4.339
81,4.322
82,4.330
83,4.363
84,4.378
85,4.440
86,4.456
87,4.553
88,4.586
89,4.601
90,4.643
91,4.876
92,4.875
93,4.871
94,4.862
95,4.906
96,5.080
97,5.044
98,5.265
99,5.250
100,5.274
101,5.440
102,5.561
103,5.545
104,5.549
105,5.649
106,5.653
107,5.681
108,5.804
109,5.845
110,5.884
111,5.916
112,5.980
113,5.996
114,6.111
115,6.868
116,6.832
117,6.827
118,7.210
119,7.318
120,7.558
121,7.614
122,7.582
123,7.653
124,8.808
125,9.131
126,9.433
127,9.582
};
\label{plot:beta1.133}
\addlegendentry{\footnotesize Graph search, greedy, $k_\mathrm{start}=64$};

\addplot [color=rot,line width = 1pt,solid]
table[col sep=comma]{
1,6.759
2,7.370
3,6.483
4,5.951
5,5.498
6,5.353
7,5.268
8,5.090
9,4.948
10,4.886
11,4.747
12,4.627
13,4.508
14,4.402
15,4.338
16,4.266
17,4.164
18,4.037
19,3.987
20,3.925
21,3.825
22,3.773
23,3.777
24,3.780
25,3.707
26,3.729
27,3.697
28,3.681
29,3.697
30,3.698
31,3.713
32,3.719
33,3.695
34,3.672
35,3.653
36,3.650
37,3.655
38,3.688
39,3.680
40,3.657
41,3.648
42,3.629
43,3.638
44,3.642
45,3.642
46,3.630
47,3.653
48,3.628
49,3.587
50,3.599
51,3.593
52,3.594
53,3.571
54,3.599
55,3.592
56,3.611
57,3.621
58,3.644
59,3.668
60,3.714
61,3.755
62,3.777
63,3.807
64,3.859
65,3.852
66,3.890
67,3.922
68,3.976
69,4.017
70,4.048
71,4.068
72,4.096
73,4.129
74,4.151
75,4.189
76,4.206
77,4.235
78,4.256
79,4.279
80,4.308
81,4.316
82,4.322
83,4.323
84,4.363
85,4.360
86,4.377
87,4.428
88,4.456
89,4.461
90,4.472
91,4.484
92,4.534
93,4.547
94,4.586
95,4.640
96,4.699
97,4.797
98,4.832
99,4.901
100,4.978
101,5.020
102,5.062
103,5.177
104,5.264
105,5.350
106,5.418
107,5.525
108,5.575
109,5.631
110,5.715
111,5.771
112,5.865
113,5.937
114,6.008
115,6.079
116,6.203
117,6.305
118,6.437
119,6.549
120,6.754
121,6.933
122,7.179
123,7.413
124,7.669
125,7.839
126,8.053
127,8.225
};
\label{plot:graphseq}
\addlegendentry{\footnotesize Graph search, $L=40$, $k_\mathrm{start}=32$};

\end{axis}
\end{tikzpicture}}
    \caption{\footnotesize Performance of rate-compatible polar code sequences with $N=128$ for \ac{BP} decoding with $N_\mathrm{it,max}=200$ iterations.}
    \label{fig:seqbp200}
\end{figure}
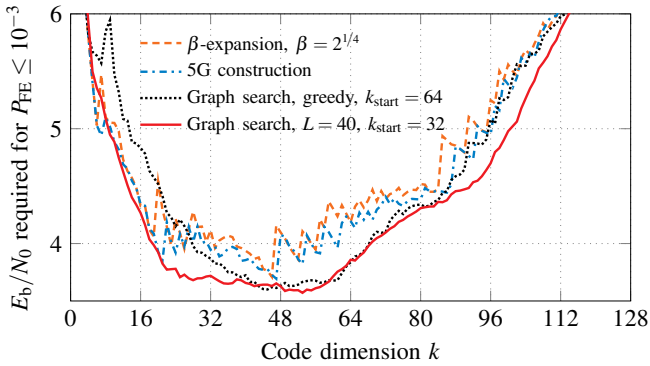
To evaluate Algorithm~\ref{alg:sequence}, we design polar codes with blocklength $N=128$ for \ac{BP} decoding with $N_\mathrm{it,max}=200$ iterations. As neither \ac{GenAlg} nor \ac{NN} based methods can optimize a rate-compatible sequence, we compare to the 5G and the $\beta$-expansion (with the standard parameter $\beta=2^{\nicefrac{1}{4}}$) sequences. To visualize the performance of the code sequence of a wide range of code rates, we plot the required $E_\mathrm{b}/N_0$ to reach an \ac{FER} of $10^{-3}$ versus the code  dimension $k$ in Fig. \ref{fig:seqbp200}. First, a greedy search ($L=1$) from $k_\mathrm{start}=64$ is performed. The sequence already outperforms both the 5G and the $\beta$-expansion sequences in the vicinity of the expansion point, however, the performance deteriorates for very high rates and in particular, low rates. Hence, we chose a lower rate expansion point $k_\mathrm{start}=32$ and also use a list $L=40$. This way, a code sequence is found that outperforms the 5G and $\beta$-expansion designs over all rates, with a maximum improvement of roughly half a dB for $k=49$. We notice that the graph search algorithm produces a sequence with much smoother transitions from one code rate to another, i.e., more predictable performance when the rate is changed, while the curves for the traditional code designs are very jagged.

\section{Conclusion}\label{sec:conc}
In this paper, we introduced a new perspective on polar code design as a search on a graph. This makes it possible to systematically optimize a single code design and also find reliability sequences for rate-compatible polar codes. To this end, we proposed two algorithms for traversing the graph and showed that they provide lower computational complexity than other Monte Carlo simulation based design methods and can result in better code designs with respect to the error-rate performance.

The proposed methods are very general and can be easily applied to other decoding algorithms such as \ac{SCAN}, \ac{BPL} and \ac{AED}. In particular, the graph can be altered such that the resulting codes follow desired properties such as the partial order of synthetic channels.

\bibliographystyle{IEEEtran}
\bibliography{references}
\end{NoHyper}
\end{document}